\documentclass[reqno,11pt]{amsart}
\usepackage{amsmath, latexsym, amsfonts, amssymb, amsthm, amscd}
\usepackage{graphics,epsf,psfrag}
\numberwithin{equation}{section}

\newtheorem{theorem}{Theorem}[section]
\newtheorem{lemma}[theorem]{Lemma}
\newtheorem{proposition}[theorem]{Proposition}

\newtheorem{rem}[theorem]{Remark}
\newtheorem{fact}[theorem]{Fact}

\renewcommand{\tilde}{\widetilde}

\newcommand{\cA}{{\ensuremath{\mathcal A}} }

\newcommand{\cB}{{\ensuremath{\mathcal B}} }

\newcommand{\bE}{{\ensuremath{\mathbf E}} }

\DeclareMathSymbol{\leqslant}{\mathalpha}{AMSa}{"36} 
\DeclareMathSymbol{\geqslant}{\mathalpha}{AMSa}{"3E} 
\DeclareMathSymbol{\eset}{\mathalpha}{AMSb}{"3F}     
\newcommand{\dd}{\,\text{\rm d}}             

\newcommand{\bbC}{{\ensuremath{\mathbb C}} }
\newcommand{\bbD}{{\ensuremath{\mathbb D}} }
\newcommand{\bbE}{{\ensuremath{\mathbb E}} }

\newcommand{\bbR}{{\ensuremath{\mathbb R}} }

\newcommand{\ga}{\alpha}
\newcommand{\gb}{\beta}
\newcommand{\gd}{\delta}
\newcommand{\gep}{\varepsilon}       
\newcommand{\gp}{\varphi}

\newcommand{\go}{\omega}

\newcommand{\gO}{\Omega}
\newcommand{\gl}{\lambda}

\makeatletter
\def\captionfont@{\footnotesize}
\def\captionheadfont@{\scshape}

\long\def\@makecaption#1#2{%
  \vspace{2mm}
  \setbox\@tempboxa\vbox{\color@setgroup
    \advance\hsize-6pc\noindent
    \captionfont@\captionheadfont@#1\@xp\@ifnotempty\@xp
        {\@cdr#2\@nil}{.\captionfont@\upshape\enspace#2}%
    \unskip\kern-6pc\par
    \global\setbox\@ne\lastbox\color@endgroup}%
  \ifhbox\@ne 
    \setbox\@ne\hbox{\unhbox\@ne\unskip\unskip\unpenalty\unkern}%
  \fi
  \ifdim\wd\@tempboxa=\z@ 
    \setbox\@ne\hbox to\columnwidth{\hss\kern-6pc\box\@ne\hss}%
  \else 
    \setbox\@ne\vbox{\unvbox\@tempboxa\parskip\z@skip
        \noindent\unhbox\@ne\advance\hsize-6pc\par}%
\fi
  \ifnum\@tempcnta<64 
    \addvspace\abovecaptionskip
    \moveright 3pc\box\@ne
  \else 
    \moveright 3pc\box\@ne
    \nobreak
    \vskip\belowcaptionskip
  \fi
\relax
}
\makeatother
\def\writefig#1 #2 #3 {\rlap{\kern #1 truecm
\raise #2 truecm \hbox{#3}}}

\newcommand{\tf}{\textsc{f}}

\begin{document}

\title[Oscillatory
critical amplitudes and the   Harris variable]{Oscillatory
critical amplitudes in hierarchical models and the tail of the Harris random variable}

\author{Ovidiu Costin}
\address{Mathematics Department, 
           The Ohio State University,
           231 W. 18th Avenue, 
           Columbus, Ohio 43210, USA}

\author{Giambattista Giacomin}
\address{
  Universit{\'e} Paris Diderot (Paris 7) and Laboratoire de Probabilit{\'e}s et Mod\`eles Al\'eatoires (CNRS),
U.F.R.                Math\'ematiques, Case 7012 (site Chevaleret)
             75205 Paris Cedex 13, France
}

\date{\today}

\begin{abstract}
Oscillatory critical amplitudes have been repeatedly observed in hierarchical models and, in the
cases that have been taken into consideration, these oscillations are so small to be hardly detectable. 
Hierarchical models are tightly related to iteration of maps and, in fact, very similar phenomena
have been repeatedly reported in many
fields of mathematics, like combinatorial evaluations and discrete branching processes. 
It is precisely in the context of branching processes with bounded off-spring
 that T. Harris, in 1948, first set forth the possibility
that the logarithm of the moment generating function  of the rescaled population size, in the super-critical regime, does not grow near infinity as a power, but it has an oscillatory prefactor. These oscillations have been observed numerically only much later and, while 
the origin is clearly tied to the discrete character of the iteration, the amplitude size 
is not so well understood. 
The purpose of this note is to reconsider the issue for hierarchical models and in what is arguably  the most elementary setting -- the pinning model -- that actually just boils down to iteration of polynomial maps (and, notably, quadratic maps). In this note we show that
the oscillatory critical amplitude  for pinning models and the oscillating pre factor connected to the Harris random variable coincide. Moreover 
we make explicit the link between these oscillatory functions  and the geometry of the Julia set of the map, making thus rigorous and quantitative some ideas set forth in \cite{cf:DIL}. 
\end{abstract}

\maketitle

\section{Introduction and main results}
\subsection{Hierarchical models, Branching processes, and the main issue}
It has been pointed out by several authors for a long time, see e.g. \cite{cf:NvL}, 
that the discrete character of the renormalization group transformation may in principle
give rise to a periodic modulation of the critical amplitude. This oscillation has been 
then observed in hierarchical models, see for example \cite{cf:DDSI,cf:DEE,cf:DIL} that deal in particular with Ising
and Potts models on diamond lattices. The modulation of the critical amplitude in these models
turns out to be very small and the nature of this phenomenon has not been fully elucidated.
Here we consider this phenomenon for a particular hierarchical model, the wetting or pinning hierarchical
 model \cite{cf:DHV}, because it is arguably the easiest set-up and it makes a direct contact with 
a vast mathematical literature: iteration of polynomial maps.

In fact the partition function of a hierarchical model with pinning potential $h\in \bbR$ and volume size $d^n$
if just $f_n(\exp(h))$, where $f_n=f\circ \ldots \circ f$ and
\begin{equation}
\label{eq:f}
f(x)\, :=\, \sum_{i=0}^d p_i x^i\, ,
\end{equation}
with $p_i\ge 0$, $\sum_{i=0}^d p_i=1$ and $p_d>0$. 
We set $w= \sum_i i p_i = f'(1)$
and $\gamma:= \log d/ \log w$: we assume
the {\sl super-criticality condition} $w>1$, so that $1$ is an unstable fixed point of $f$
and the sequence $\{f_n(\exp(h))\}_{n=0,1,\ldots}$ increases to infinity. Actually
it is straightforward to see that the increase is super-exponentially fast and just
a little more work leads to the existence of the free energy density of the model
\begin{equation}
\label{eq:freeenergy}
\tf(h)\, :=\, \lim_{n \to \infty} \frac1{d^n} \log f_n(\exp(h))\, .
\end{equation} 
We refer to \cite[Appendix]{cf:GLT} for the existence of this limit as well as for 
various properties of $\tf$, such as the fact that $\tf$ is a convex  non-decreasing function.
As a matter of fact, it is immediate to see that
$\tf(h)=0$ for $h \le 0$, while $\tf(h)>0$ for $h>0$. The origin is therefore necessarily a critical point
and it is actually
associated to a {\sl localization transition} (see \cite{cf:DHV,cf:GLT} and references therein for details on the statistical mechanics context, about which we are particularly concise here).

One is then particularly interested in the free energy critical behavior,
that is how $\tf(h)$ behaves near $0$, which, in this case, of course reduces to considering $h \searrow 0$. What is argued in the aforementioned  physical literature  is that the expected critical behavior must be of the form
\begin{equation}
\label{eq:Fp}
\tf(h) \stackrel{h \searrow 0} \sim h^{\gamma}  A ( \log h)\, ,
\end{equation}
where the amplitude $A:\bbR \to (0, \infty)$  is $\log w$-periodic, which
is of course compatible with $A(\cdot)$ being constant.

\medskip
\begin{rem}\rm
We point out that the hierarchical
model in \cite{cf:DHV} has $d=2$, $p_0=(B-1)/B$, $p_1=0$
and $p_2=1/B$, with $B \in (1,2)$ (to be precise  in \cite{cf:DHV} the {\sl dual} model
with  $h$ replaced by $\log(B-1)+h$ and $B\in (2,\infty)$ is considered,
see \cite{cf:GLT} for the equivalence of these two models).   
The restriction to  $d=2$ is just for the sake of simplicity, but  the choice 
of $p_1=0$
reflects a symmetry of the model that has no particular impact on the issue we tackle in this note: 
\cite{cf:DHV,cf:GLT} deal with {\sl disordered} hierarchical pinning and in the presence 
of disorder the fact that $p_1>0$
leads to  a new phenomenon (but this is not not the case in absence of disorder). 
\end{rem}  

\medskip

In \cite{cf:DDSI} and with a substantially more detailed analysis in \cite{cf:DIL} the authors 
tackle the issue of establishing whether  $A(\cdot)$ is trivial or not and of understanding the origin and size of the fluctuations. In \cite{cf:DIL} an argument is presented, in the Ising and Potts models framework, that is expected to capture the size of the critical amplitude oscillations, but it yields an
amplitude  that is smaller than the true one, obtained numerically, by several orders of magnitude. The authors argue heuristically about the reason of such a  mismatch between their arguments and numerics and they point out the  role of the geometry of the
Julia set of the map associated to this model. In this direction they actually provide 
a relatively precise estimate of the oscillation amplitude by exploiting empirical estimates on the Julia set, even if they admit that the precision of their computation is surprising and
{\sl presumably merely to be considered as a lucky circumstance} \cite[p.~123]{cf:DIL}.
Clarifying the relation between Julia set and oscillations is one of the main aims of this note.  

An important observation at this point is that there is a probabilistic representation of the 
partition function of the hierarchical pinning model \cite{cf:GLTcpam}. Consider in fact
a random branching process, or Galton-Watson process, starting from one individual 
 at time zero and with offspring distribution determined by the $p_j$'s weights. 
Therefore if one focuses on the number of individuals $W_n$ that are present 
at time $n$, then, conditionally on $W_n=k$, $W_{n+1}$ is the sum of $k$ independent 
and identically distributed random variables for which the probability of being equal to $j$ is $p_j$. 
One directly verifies that 
\begin{equation}
\label{eq:dv}
f_n(\exp(h))\, =\, \bbE \left[ \exp( hW_n)\right]\, .
\end{equation}

\subsection{Branching processes and  the Harris random variable}

It is a classical result that $W_n/ w ^n$ converges almost surely to a non-degenerate limit random variable $W$ (with a mass at zero if $p_0>0$). Moreover for every $s\in \bbR$
\begin{equation}
\label{eq:toexp}
\lim_{n \to \infty} \bE \left[ \exp(s W_n/ w ^n)\right]\, = \, \bE\left[\exp(s W) \right]\, =: \, \psi(s)\, ,
\end{equation}
and $\psi (\cdot)$ extends to the whole complex plane as an entire function. 
In his seminal paper \cite{cf:Harris}, T.~E.~Harris pointed out (among several other facts)
that
\begin{equation}
\label{eq:Harris}
\log\psi(s) \stackrel{s \to \infty }\sim  s^\gamma L(\log s)\, ,
\end{equation}
where $L: \bbR\to (0,\infty)$ -- the {\sl Harris function} -- is continuous and  
$\log w$-periodic.
 Harris was unable to show that $L(\cdot)$ is not constant, even if
he was able to compute numerically the value of $L(s)$ for $d=2$, $p_1=0.4$
and $p_2=0.6$ up to six decimal digits, a remarkable
achievement considering the date at which the paper was published. 
Later it became clear that $L(\cdot)$ does oscillate 
 and  that the amplitude of the oscillation is extremely small
with respect to its {\sl average} value (see in particular \cite{cf:BN}, but also
\cite{cf:BB1,cf:BB2,cf:jones}).  A full understanding of this near-constant behavior is however still elusive. 
  Actually analogous phenomena were recorded also in other 
mathematical fields like combinatorial enumeration, spectral properties of transition operators 
on fractals and more (see e.g. \cite{cf:Odlyzko,cf:GrabWoess,cf:Teufl,cf:Hambly}). Ultimately, this is not
surprising because all these problems boil down to studying iterations of a map: in the Galton-Watson case for example \eqref{eq:dv} tells us that 
 the generating function of the law of $W_n$ is precisely $f_n$.

Certainly it has not escaped the reader that the qualitative properties of $L(\cdot)$
coincide with the (expected) qualitative properties of the critical amplitude $A(\cdot)$.
However the  quantitative connection between $L(\cdot)$ and $A(\cdot)$, beyond the common
 period $w $, is a priori not  clear. In statistical mechanics terms, the Harris function $L(\cdot)$
 emerges from the 
limit of the partition function in a particular vanishing limit of the pinning parameter $s/ w ^{n}$:
$n\to \infty$ and then $s \to \infty$. $A(\cdot)$ comes also out of a limit of vanishing pinning
parameter, but in this case $n \to \infty$ is taken at fixed pinning parameter $h>0$,
only the leading Laplace asymptotic term is kept, and then $h$ is sent to zero. 

Nevertheless,  $L(\cdot)$ and $A(\cdot)$ coincide: 
\medskip

\begin{proposition}
\label{th:qualit}
Given $f(\cdot)$ as above, the asymptotic relation \eqref{eq:Fp} holds with $A(\cdot)$
a  $\log w$-periodic analytic function. Moreover $L(\cdot)=A(\cdot)$
-- from now on, they will be denoted by $\gO(\cdot)$ -- and, if we denote
 $c_n = \int_0^1 \exp(i n x) \gO( x \log w) \dd x$ 
the Fourier coefficients of $\gO(\cdot \log w)$,
there exists a positive constant
$\gep$ such that $\vert c_n \vert \le \exp(-\gep n)$ for $n$ sufficiently large.
\end{proposition}
\medskip

The notation we use does not highlight the dependence of $\gO(\cdot)$ on $f(\cdot)$, but we stress that
Proposition~\ref{th:qualit} says that given $f$ one obtains a function
$\gO$.

Proposition~\ref{th:qualit} can be proven in a rather direct way by exploiting a number
of relations that one can find in the large literature devoted to the subject, but we have been 
unable to find the statement in this literature.
The proof is  in Section~\ref{sec:proofs} but let us anticipate one of the main tools that is going to be of help for the sequel of the introduction: for every $x>0$
\begin{equation}
\label{eq:F.1}
\tf(\log x)\, =\, \frac{\log  p_d}{d-1}+ \log x+ 
\sum_{i=0 }^\infty  d^{-(i+1)}{Q(f_i(x))}\,, 
\end{equation}
with $Q(x):=\log( f(x)x^{-d}/p_d)$. This follows by observing that, if $\tf_0(y)=y$ and $\tf_n(\log x):=d^{-n}\log f_{n}(x)$ for $n=1,2, \ldots$ 
we have   $\tf_n(\log x)=d^{-n}\log p_d  + d^{-n} Q(f_n(x))+ \tf_{n-1}(\log x)$ \cite{cf:Harris}.

To go beyond Proposition~\ref{th:qualit} we restrict to the $d=2$ case. This restriction 
is made because we want to exploit directly the results in 
 \cite{cf:CHadv,cf:CH,cf:CK}, that develop only the case $d=2$.
 It is certainly possible to generalize these works, but this
 would not add much to the purpose of this note at the expense of rather lengthy arguments. 
 
 \subsection{Julia set, B\"ottcher function and oscillations}
 \label{sec:Julia_etc}
 If we set $\cB(x):=\exp(\tf(\log x))$ we directly verify that   $\cB$ 
 solves 
the {\sl B\"ottcher equation} $\cB(f(x))\, =\, \cB(x)^d$. This
can be seen directly from \eqref{eq:F.1}, that is $\cB$ conjugates $f$ and the monomial map $x 
\mapsto x^d$. $\cB$ is increasing on $[1, \infty)$ and it is real analytic on $(1, \infty)$ (see the proof of Proposition~\ref{th:qualit}). 
 We can therefore set  $\cA:= \cB^{-1}$ and $\cA$ is analytic too. 
 
The central point of our approach is the following theorem, in which we use the notation
$\bbD_r:= \{z \in \bbC:\, \vert z \vert <r\}$. 
In what follows $\log(\cdot)$ is the natural logarithm with the choice of the negative semi-axis 
as branch cut: $\log (r \exp(i \theta))=\log r + i\theta$, for $r>0$ and $\theta \in (-\pi, \pi]$.
Let us recall \cite{cf:Milnor} that the filled Julia set $K=K(f)$ of the polynomial map $f$ is the
 set $\{z\in \bbC:\, \sup_n \vert f_n(z)\vert < \infty\}$ and that the Julia set $J=J(f)$ is the boundary
 of the filled Julia set. 

It is immediate to see that  $J=\{ z\in \bbC : \, \vert z \vert =1\}$
if $f(x)=x^d$, that is if $p_d=1$ (recall \eqref{eq:f}), but $J$ is substantially more complicated 
if $p_d<1$. However 
 note also that, for $d=2$, $f$ is conjugated to the map $g(x)=x^2+c$,
$c=(1-(p_2-p_0)^2)/4$, via the affine transformation
$h(x)=x/p_2+ (1-p_0-p_2)/2p_2$, that is $g= h^{-1}\circ f \circ h$ and that
$c\in (0,1/4)$: this implies  
that the Julia set is a simple closed curve
(see for example \cite[Ch. 3, Prop.~6.2]{cf:Dev}).   

Set $d=2$ and $0<p_2<1$. 
\medskip

\begin{theorem}
\label{th:Bottcher} 
The function $\cA$  extends analytically to  
$\bbC \setminus \overline{\bbD}_1$ and to $\partial \bbD_1$ as a continuous function. 
Moreover there exist an entire function $g$, with
$g(0)=1$ and $g'(0)>0$, and a $\log 2$-periodic  function $\go(\cdot)$, which is non-trivial (i.e. non-constant), positive and 
analytic on the strip $\{z:\, \vert \Im(z)\vert<\pi /2\}$, such that  
for $\vert z\vert >1$  
we have
\begin{equation}
\label{eq:Bottcher1} 
\cA (z)\, =\, g\left( (\log z)^{1/\gamma} \go(\log(\log z)) \right)\, .
\end{equation}
In addition:
\begin{enumerate}
\item the Julia set of $f$   coincides with 
$\{ \cA (\exp( i\pi  t):\, t \in (-1, 1] \}$.
\item the function $x\mapsto x^{1/\gamma} \go (\log(x))$, from the positive semi-axis to itself,  is invertible and its inverse is $x \mapsto x^\gamma \ga(\log (x))$,
where $\ga$ is $\log w$-periodic and $\gO(\cdot)= \mathtt c ^\gamma \ga (\cdot +\log(\mathtt c))$,
with $\mathtt c:= p_2/(2-w)$. Moreover $\ga(\cdot)$, and therefore $\gO(\cdot)$,
 is analytic in the strip 
$\{z:\, \vert \Im(z)\vert < \pi/(2 \gamma)\}$, which implies that
$\int_0^{\log w} \exp(2\pi inx/ \log w)\gO(x)\dd x =O(\exp(-n\pi ^2 / \gamma'))$, for every $\gamma' > \gamma$ and $n \to \infty$.
\end{enumerate}
\end{theorem}

\medskip

To keep the statement simple the explicit  series expansion for $g(\cdot)$
is postponed to Section~\ref{sec:quantit}, starting from \eqref{eq:gtog}.
In Section~\ref{sec:quantit} one can also find  an explicit construction for $\go(\cdot)$.
What we want to emphasize with Theorem~\ref{th:Bottcher} is the quantitative and explicit relation between oscillations
and geometry of the Julia set. 

\medskip

\begin{rem}\rm 
A more general but very implicit relation between Julia set and oscillations can be established
 by using the notion of Green's function of the  monic polynomial  map $\tilde f$
\cite[P.~100-101]{cf:Milnor}, where {\sl monic} means that $\tilde f(x)=x^d+O(x^{d-1})$. The Green's function $u(z):= \lim_{n\to \infty} d^{-n} \log \vert \tilde f_n(z)\vert$
is well defined and in $C^0( \bbC ; \bbR)$. It is actually harmonic ($(\partial_x^2 + \partial^2_y) u(x+iy)=0$) 
 except on  the Julia set $J(\tilde f)$ and it is of course identically zero on the filled Julia set $K(\tilde f)$. 
Moreover $u$ can be alternatively defined as the unique continuous function
that vanishes on $K(\tilde f)$, which is harmonic in $J(\tilde f)^\complement$ and  such that $\lim_{z \to \infty} u(z) - \log z=0$. Note the remarkable fact that $u(\cdot)$ depends on
$\tilde f(\cdot)$ only through $J(\tilde f)$. If now we observe that $f$ is conjugated 
to a monic $\tilde f$ via the linear map $z \mapsto  cz$, $c:=p_d^{-1/(d-1)}$, we readily see that
$J$ is $c J(\tilde f)$ and $\tf(\log x)= u(x/c)$ for $x\in (0, \infty)$. Therefore  the free energy   is determined as unique the solution of a Dirichlet problem
once $J(f)$ (and the blowing up factor  $c$) are given. As a consequence, the
oscillatory function
$\gO(\cdot)$ is directly connected to the Julia set via the solution of this Dirichlet problem. 
\end{rem}

\medskip

Theorem~\ref{th:Bottcher}  asserts that $\gO(\cdot)$ is not a constant: estimating the size of the oscillations
is a challenging task. However, one does have 
 {\sl explicit} characterizations of $\go$  and $\gO$
that can be exploited  to get precise numerical and also {\sl computer-assisted} estimates on the Fourier 
coefficients of $\go$, with explicit error bounds. We address this issue in Section~\ref{sec:quantit}.



\begin{figure}[hlt]
\begin{center}
\leavevmode
\epsfxsize =12 cm
\epsfbox{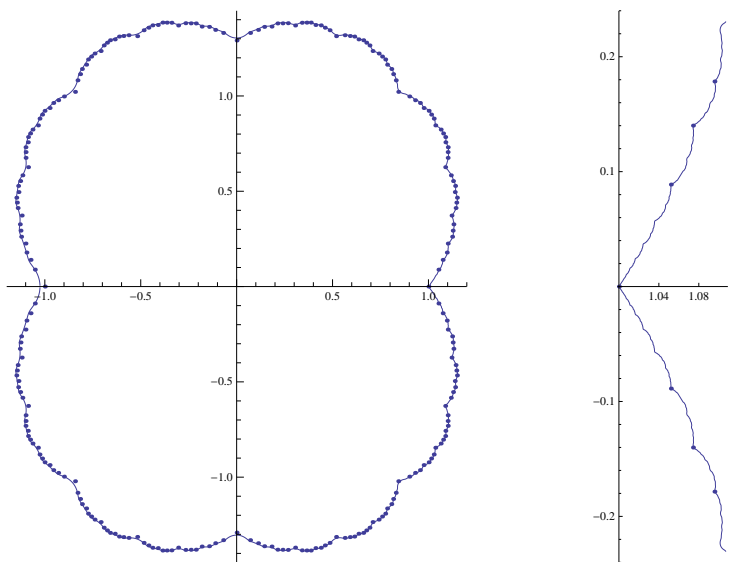}
\end{center}
\caption{The Julia set of $f(x)=1/4 + 3x^2/4$ plotted by using 
$J=\{ \cA (\exp( i \pi t):\, t \in (-1, 1] \}$
 and
\eqref{eq:Bottcher1}, after having determined $\cA(\cdot)$ as explained in Remark~\ref{rem:add}: a priori
the method determines $\cA(\cdot)$ with precision in a neighborhood of $1$ in the complex plane, as it is confirmed by
the figure on the right, but the precision is quite satisfactory also far from $1$. 
We have superposed to this curve 
 the points obtained
by  iterating (seven times)  the pre-image map $f^{-1}$, starting from the unstable fixed point $1$.}
\label{fig:1}
\end{figure}

\begin{figure}[hlt]
\begin{center}
\leavevmode
\epsfxsize =14 cm
\epsfbox{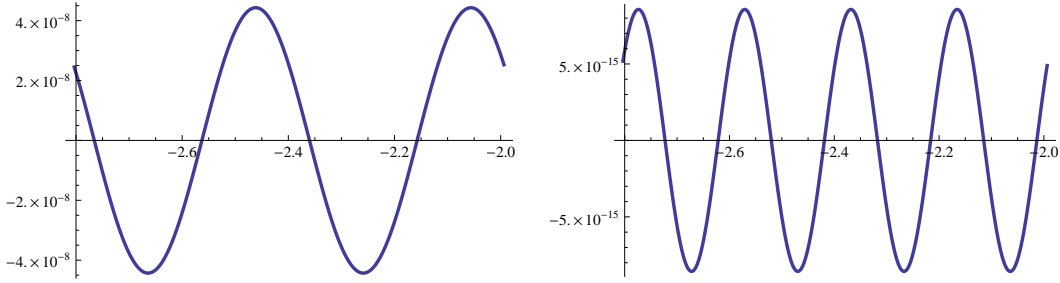}
\end{center}
\caption{On the left the plot of $\gO(\cdot)- \overline{\gO}$, with $\overline{\gO}:= \frac 1{\log w}\int_0^{\log w} \gO(x)\dd x$ obtained by using  $f(x)=1/4 + 3x^2/4$, that is
the same map as in Figure~\ref{fig:1}. In particular the period $\log w$ is $0.40546\ldots$ and 
we have $\overline{\gO} = 1.33381\ldots$
and the oscillation is $8.86\ldots \times 10^{-8}$. On the right we plot $\gO(\cdot)- \overline{\gO}- c_1 \sin(2\pi (\cdot- x_1)/\log w)$, with $c_1$  the first Fourier coefficient and $x_1$ the associated shift. This second plot makes
clear 
the  nearly sinusoidal character of $\gO(\cdot)$ as well as the fact that the next Fourier coefficient is even much smaller: in fact the plot on the right is again nearly sinusoidal.}
\label{fig:2}
\end{figure}

\section{Proofs}
\label{sec:proofs}
\medskip

\subsection{Proof is Proposition~\ref{th:qualit}}
We have the following representation of the Harris function \cite{cf:Harris,cf:BN}
$L(\cdot)$ given in \eqref{eq:Harris}: for $s>0$
\begin{equation}
\label{eq:Hrep}
L(\log s)\, =\, s^{-\gamma}\tf(\log \psi(s))\,  .
\end{equation}
It is worth recalling that \eqref{eq:Hrep} follows 
by using the {\sl Poincar\'e relation} $\psi( ws)=f(\psi(s))$, which is an immediate consequence of
\eqref{eq:toexp},  and of   $\tf(\log f(s))= d \tf ( \log(s))$ which is  equivalent to the B\"ottcher functional relation (cf. \S~\ref{sec:Julia_etc}). In fact
these two relations imply that if $\tilde L(\cdot)$ is the right-hand side of \eqref{eq:Hrep} then we have
\begin{equation}
\tilde L(\log(ws))\, =\, \frac{\tf(\log(\psi(w  s)))}{w ^\gamma s^\gamma}\, =\, \frac{\tf ( \log f (\psi( s)))}{w ^\gamma s^\gamma} \, =\, \frac{\tf (\log (s))}{s^\gamma}\, =\, \tilde L(\log s)\, .
\end{equation}
But \eqref{eq:F.1} directly implies that  $\tf(h) \stackrel{h \nearrow \infty}\sim h$ and so
\begin{equation}
\tilde L (\log s)\,=\, \frac{\tf(\log(\psi(w^n  s)))}{\log(\psi(w^n  s))}\frac{\log(\psi(w^n  s))}{(w ^{n} s)^\gamma} \stackrel{n \to \infty}\sim L(n \log w + \log s)\, =\, 
 L(\log s)\, , 
\end{equation}
where in taking the limit we have also used  \eqref{eq:Harris}. Therefore \eqref{eq:Hrep}
is proven.

Going back to the main argument, 
$\psi:[0, \infty) \to [1, \infty)$ is increasing and $\psi(s)=1+s+o(s)$, for $s$ small,
so $\psi^{-1}(\exp(h)) \sim h$ for $h$ small. Therefore, since $L (\cdot)$ is continuous (and periodic),  we obtain from \eqref{eq:Hrep}
\begin{equation}
\tf( h ) \stackrel{h \searrow 0} \sim 
h^\gamma L( \log h)\, ,
\end{equation}
which is \eqref{eq:Fp} with $A(\cdot)=L(\cdot)$.

We are therefore left with the regularity properties of $A(\cdot)=L(\cdot)$
which we now call $\gO(\cdot)$. For this we first claim that
 that
$x \mapsto \tf (\log(x))$ extends to an analytic function in a cone
$C_\gd:=\{z:\, \vert \text{arg}(z-1)\vert < \gd$ and $\Re(z-1)>0\}$, for a $\gd>0$. 

The claim follows from by \eqref{eq:F.1}  because one 
can find $c>1$ such that   both $\vert (f(z)z^{-d}/p_d)-1 \vert \le 1/2$ and
$\vert f(z) \vert \ge c$ for $\vert z \vert \ge c$. Therefore no singularity comes from
the series in  \eqref{eq:F.1} if $x\in C_\gd$, if $\gd$ is sufficiently small, because 
by elementary estimates we see that the argument of $f_n(x)$ remains bounded 
by a constant smaller than $\pi$ for $n \le \min\{n: \, \vert f_n(x) \vert \ge c\}$.

Let us now look at  the formula for $\gO(\cdot)$ in \eqref{eq:Hrep}
with $1+s \in C_{\gd'}$ for $0< \gd'< \gd (< \pi)$:
by periodicity it suffices to consider 
$\Re (s) $ small, which guarantees that the (entire) function $\psi(s)=1+s+O(s^2)$
takes values in the  cone $C_\gd$ and therefore $\gO(\cdot)$ is analytic in the 
strip $\{ z : \, \vert \Im (z)\vert < \gd\}$, which directly implies 
that the Fourier coefficients of $\gO(\cdot)$ are smaller than
$\exp(2\pi \gd' / \log w)$, for every $\gd'< \gd$ and $n$ sufficiently large.
This completes the proof of Proposition~\ref{th:qualit}. 
\qed

\medskip

The argument we have just presented effectively  uses the analyticity of $\tf(\log(z))$
for $z$ in a truncated cone, that is
for $z \in C_\gd$ and $\Re (z-1)$ small. In order to get a better (in fact, optimal)  estimate on the
decay of the  Fourier coefficients a precise knowledge of the Julia set is needed,
in the sense that it is necessary for the truncated cone to be in the
complement of the filled Julia set. This of course guarantees that $f_n(z)\to \infty$, but this a priori is not sufficient because one has to
ensure also that the absolute value of the argument of $f(z_n)z_n^{-d}/p_d$, $z_n
:= f_n(z)$, does not go beyond 
$\pi$ for every $z=z_0$ in the truncated cone (recall \eqref{eq:F.1}). We therefore restrict to $d=2$
and attack the problem from a somewhat different angle.

\subsection{Proof of Theorem~\ref{th:Bottcher}}
For the map $f$ we consider $\infty$ is a super-attracting fixed point
(see \cite[p.~44-45]{cf:Milnor}). It is practical and customary
to map the fixed point at infinity to a fixed point at zero by conjugation.
Let us make this explicit  for $d=2$, so that
$f(x) = p_0+ p_1 x +p_2 x^2$
and  $w  = p_1+2p_2\in (1,2)$.
The fixed points of $f$ are $1$ (unstable)
and $p_0/p_2<1$ (stable).
The affine transformation $x \mapsto l(x)$
\begin{equation}
l(x)\, :=\, \frac{p_0}{2-w }-x\frac{p_2}{2-w }\, ,
\end{equation}
sends $f(\cdot)$ into the standard logistic map:
\begin{equation}
l(f(l^{-1}(z)))\, =\, \gl z(1-z)\, , \ \ \ \ \gl\, :=\, 2-w  \in (0,1)\, . 
\end{equation}
The fixed points are mapped to $1-1/\gl$ (unstable) and $0$ (stable). We then map $z$ to $-1/z$, so in the end we use the transformation 
$q(x):= -1/ l(x)$ ($q: \bbR \setminus \{p_0/p_2\} \longrightarrow \bbR \setminus \{0\}$)
to get to $\mathtt{f}(y)= q(f(q^{-1}(y)))$ and the new iteration
\begin{equation}
\label{eq:y}
y_{n+1}\, =\,  \mathtt{f}(y_n) \, =\, \frac{y_n^2}{\gl(1+y_n)}\, ,
\end{equation}
and the unstable fixed point is now $\gl/(1-\gl)$, while  $0$ is stable (in fact, super-attracting).  
For later use, 
it is of help to
 note that $q(1)=\gl/(1-\gl)$ and that
 \begin{equation}
 \label{eq:record}
 \gd(h)\, :=\, \frac{\gl}{1-\gl}-q(\exp(h)) \stackrel{h \searrow 0}\sim c_\gl h \, ,
 \ \ \ \ \ \text{ with} \  c_\gl:= \frac{p_2 \gl} {(1-\gl)^2}\, .
 \end{equation}
 
 \medskip
 
 We are going to use the fact that there exists 
 a unique function $\mathtt G$, from $\bbD_1$ to the interior of $K(\mathtt f )$, which
 is analytic and invertible, satisfying $\mathtt G (0)=0$ and  $\mathtt G' (0)=\gl$ and
\begin{equation}
\label{eq:Bottcher3}
\mathtt G^{-1} \circ \mathtt f \circ \mathtt G (z)\, =\, z^2\, .
\end{equation}
Actually, the existence and uniqueness of such a map in a small disk around the origin 
is a general result (B\"ottcher Theorem \cite[\S~9]{cf:Milnor}). 
 $\mathtt G^{-1}  $ is usually called  B\"ottcher function for $\mathtt{f}$ and
 it is uniquely determined once we require it to be analytic near zero, with
 $\mathtt G ^{-1} (0)=0$ and $(\mathtt G ^{-1})' (0)=1/\gl$ (this is a classical result: see  \cite{cf:CH}
 and references therein). The fact  that $\mathtt G$ can be extended to the whole
 unit disk depends on the details of the map and, in our case, on the fact that $\gl \in (0,1)$
 \cite[p.~1312]{cf:CH}.  

The argument that follows is based on 
 \cite[Theorem~1]{cf:CH} which gives an expression for 
$\gp(\cdot):= 1/ \mathtt G(\cdot)$: there exists $\mathtt g(\cdot)$  entire, with
$\mathtt g(0)=0$ and $\mathtt g'(0)=1$, and a non-trivial (non-constant!) periodic function $\go(\cdot)$ of period $\log 2$
(analytic on the strip $\{z:\, \vert \Im(z)\vert<\pi /2\}$) such that  
\begin{equation}
\label{eq:CH-rep}
\gp (z)\, =\, \frac{1-\gl} \gl + \mathtt g\left( (-\log z)^{1/\gamma} \go(\log(-\log z)) \right)\, ,
\end{equation}
and we recall that $\gamma= \log 2/ \log w>1$. 
The properties of $\mathtt G(\cdot)$ directly imply that $\gp(\cdot)$ is analytic in the punctured unit disc 
and that it is a conformal (and invertible) map. Actually,
$\gp(\cdot)$ extends by continuity to  $\partial \bbD_1$
 \cite[p.~1313]{cf:CH}. Moreover $\mathtt g(\cdot)$ is invertible in the range of its argument 
 in \eqref{eq:CH-rep}.

Let us point out that \eqref{eq:CH-rep} implies \eqref{eq:Bottcher1}. For this
let us us go back to the \eqref{eq:F.1}, with which we have defined $\cB$ and then $\cA$,
see \S~\ref{sec:Julia_etc}. 
As we have already remarked in the proof of Proposition~\ref{th:qualit}, $\vert f_n(z)\vert \ge  \vert z \vert$ for $\vert z \vert $ sufficiently
large and this directly implies the analyticity of $\cB (\cdot)$ in a neighborhood of infinity,
as well as the existence of it inverse $\cA(\cdot)$ in the same neighborhood, since 
$\cB(z) \sim p_2 z$ for $\vert z \vert \to \infty$ (we remark that $\cB(\cdot)$ coincides
with the function $\widehat \psi $ in \cite[p.~98]{cf:Milnor}). 
Going (backward) through the conjugation that we have performed we see that
$1/ \cB (q^{-1}(z))\sim z/\gl$  for $\vert z \vert $ large and that  $z \mapsto 1/ \cB (q^{-1}(z))$
solves the B\"ottcher equation (in the sense that it has the same property 
as $\mathtt G^{-1}$ in \eqref{eq:Bottcher3}). It must therefore coincide with $\mathtt G^{-1}$ and  this extends the
domain of analyticity, in fact bi-analyticity, of $\cA(\cdot)$. 
Again, by going carefully through the backward conjugation we see that  $g(\cdot)= 1+ (\gl/p_2) \mathtt g (\cdot)$, so \eqref{eq:Bottcher1} is proven. The  fact that $\cA(\cdot)$ extends as a continuous function to $\partial \bbD_1$ follows from the analogous property for $\gp(\cdot)$.
\medskip

Point (1) of Theorem~\ref{th:Bottcher}  follows from 
the general theory \cite[Theorem~9.5]{cf:Milnor}. In fact
the B\"ottcher function $\cB(\cdot)$ maps (in a bi-holomorphic fashion) 
$K(f)^\complement$ to $\bbD_1^\complement$ and  $\cA(\partial \bbD_1)= J(f)$.

\medskip

Let us move to point (2)
 and let us start by remarking that
\begin{equation}
\label{eq:HM1}
\tf(h)\, =\, \lim_{n \to \infty} \frac 1{2^n}
 \log f_n(\exp(h))\, =\, -\lim_{n \to \infty} \frac 1{2^n}
 \log \mathtt f _n \left( 
q(\exp(h))\right)\, .
\end{equation}  
Therefore, in view of \eqref{eq:record}
we need to control $\mathtt f _n ( (\gl/(1-\gl)) -\gd)$ for $n\to \infty$ and then $\gd$ small. 
But \eqref{eq:Bottcher3} tells us that 
for every $n$
\begin{equation}
\mathtt f _n (\mathtt G (x))\, =\, \mathtt G \left(x^{2^n}\right)\,.
\end{equation} 
and  therefore 
\begin{equation}
\label{eq:fB}
\mathtt f _n \left( \frac{\gl}{1-\gl} -\gd \right)\, =\, \mathtt G \left( \mathtt G^{-1}\left(\frac{\gl}{1-\gl} -\gd
\right)^{2^n}\right)\,.
\end{equation} 
It is practical to set (in particular  for $\gd >0$)
\begin{equation}
\label{eq:ell}
\ell (\gd)\, :=\, -\log  \mathtt G^{-1}\left(\frac{\gl}{1-\gl} -\gd
\right)\, .
\end{equation}

We claim the following:
\medskip

\begin{lemma}
\label{th:ell}
The map $x \mapsto x^{1/\gamma} \go (\log x)$ is a bijection from $(0,\infty)$ to itself and its inverse can
be written as $x \mapsto x^\gamma \ga (\log(x))$, with $\ga(\cdot)$ $\log w$-periodic. Moreover
$\ga(\cdot)$  is analytic in the strip $\{z:\, \Im (z) < \pi/(2\gamma)\}$. Finally for $c=((1-\gl)/\gl)^2$ we have
\begin{equation}
\ell (\gd) \stackrel{\gd \searrow 0} \sim 
(\gd c)^\gamma \ga \left( \log (\gd c) \right)\, .
\end{equation}
\end{lemma}
\medskip

\noindent
{\it Proof.}
We use the properties of $\gp(\cdot)$ (or $\mathtt G (\cdot)$) and \eqref{eq:CH-rep}.
Therefore we see that if we set
 $z=\exp(-\ell (\gd))$ we have
\begin{equation}
\label{eq:2.16}
\gp\left( \exp\left( -\ell (\gd)\right)\right)\, =\, \frac{1-\gl}{\gl -(1-\gl) \gd}\, =:\,
\frac{1-\gl}{\gl}+v(\gd)\, ,
\end{equation}
and $v(\gd)\sim c \,\gd$, $c$ given in the statement. Thus we have to solve
\begin{equation}
\label{eq:tobesolved}
\ell(\gd)^{1/\gamma} \go \left(\log(\ell(\gd))\right) \, =\, \mathtt g^{-1}
\left( v(\gd)\right) \Big(\stackrel{\gd \searrow 0} \sim  c\,  \gd\Big) \, ,
\end{equation}
at least for $\gd$ small. We see therefore that it is a matter of inverting $x\mapsto x^{1/\gamma} \go(\log x)=: \gb(x)$. First, observe that 
$\gb: (0, \infty) \to (0, \infty)$ is invertible, because $\gp$ is invertible. Moreover $\gb(\cdot)$ is (real) analytic, hence $\gb^{-1}(\cdot)$ is too. Then
set $\ga (\log y):= \gb^{-1}(y)/y^{\gamma}$,
so that $\ga(\cdot)$ is real analytic and 
\begin{equation}
\label{eq:forinv}
\ga \left(\frac 1\gamma \log x+ \log \go( \log x)\right) \,= \, \frac1{\go^{\gamma}(\log x)} .
\end{equation}
 From \eqref{eq:forinv} we directly see that, since $\go (\cdot)$ is $\log 2$-periodic, 
 $\ga(\cdot)$ is $(\gamma^{-1}\log 2)$-periodic (that, is $\log w $-periodic) and, since $\go(\cdot)$ is not constant, $\ga(\cdot)$ is not a constant either .
Therefore  the statement is proven, except for the domain of analyticity. To this end note that
the argument above was given by restricting to the real axis, but we do know that
$\go(\cdot)$ is analytic on the symmetric strip of half-width $\pi/2$
and therefore $\gb(\cdot)$ is analytic in the positive half-plane 
$\{z:\, \Re (z)>0\}$: all we need to know 
is the fact that it is invertible in this domain. 
But since $\gp(\cdot)$ is defined and invertible in the punctured unit disk,  $z \mapsto \gp(\exp(-z))$ is defined and invertible in $\{z:\, \Re (z)>0\}$ 
and, thanks to \eqref{eq:2.16} and \eqref{eq:tobesolved},  we see that also $\gb(\cdot)$ is invertible, at least if we restrict to 
the  truncated cone of the points $z$ such that
$\vert \text{Arg}(z) \vert \le \pi/(1 + \gep)$, for any  choice of $\gep>0$,
and $\Re(z)>0$ sufficiently small.  Once again, since $\ga(\cdot)$ is periodic, 
analyticity on $\{z:\,  \vert \Im(z)\vert < \pi/(2\gamma(1+\gep))$ and 
$\Re(z)<-C\}$, for $\gep>0$ arbitrary and $C=C(\gep)$ large, implies analyticity  on the whole strip $\{z:\, \vert \Im(z)\vert < \pi/(2\gamma)\}$. 
\qed

\medskip

Let us complete the proof of (2)  and for this 
let us go back to
\eqref{eq:HM1} and use \eqref{eq:record} and \eqref{eq:fB}
to see that 
\begin{equation}
\tf(h)\, =\, 
-\lim_{n \to \infty} \frac 1{2^n} \log \mathtt G \left( \mathtt G^{-1}\left(\frac{\gl}{1-\gl} -\gd(h)
\right)^{2^n}\right)
\end{equation}
A direct consequence of Lemma~\ref{th:ell} and of  $\lim_{h  \searrow 0}\gd
(h)=0$ is that  $\mathtt G^{-1}(({\gl}/{1-\gl}) -\gd(h)
)<1$ for $h $ sufficiently small and therefore,
since $\mathtt G (0)=0$ and $\mathtt G '(0)>0$, for such values of $h$, we have
\begin{equation}
\tf(h)\, =\, 
-\lim_{n \to \infty} \frac 1{2^n} \log  \mathtt G^{-1}\left(\frac{\gl}{1-\gl} -\gd(h)
\right)^{2^n}\, = \, -\log G^{-1}\left(\frac{\gl}{1-\gl} -\gd(h)
\right)\, =\, \ell (\gd(h))\, .
\end{equation}
We now recall once again \eqref{eq:record} and apply Lemma~\ref{th:ell} to obtain
\begin{equation}
\tf(h) \stackrel{h \searrow 0}\sim 
(c_\gl c\,  h) ^{\gamma } \ga\left(
\log (c_\gl c\,  h)  \right)\, ,
\end{equation}
and we compute $c_\gl c=  p_2/\gl=\mathtt c$, from which we identify $\gO(\cdot)$
in terms of $\ga(\cdot)$ and $\mathtt c$.
\qed

\medskip

\medskip

\begin{rem}\label{rem:BPcase}\rm
Of course one can upgrade the proof we just completed to include
Proposition~\ref{th:qualit}, that is to deal with the asymptotic behavior 
of $\psi(\cdot)$. This is straightforward, albeit a bit lengthy: we sketch the main steps
and make some comments.  
First of all we have
\begin{equation}
\label{eq:BP1}
\psi(s)= \lim_{n \to \infty} q^{-1}
\left( \mathtt{f}_n \left(q(\exp(s/w ^n))\right)\right)\, , 
\end{equation}
and we notice the analogy with
\eqref{eq:fB} and we write
\begin{equation}
\label{eq:BP4}
 \mathtt{f}_n \left(q(\exp(s/w ^n))\right)\, =\,  \mathtt G \left( \mathtt G^{-1}\left(\frac{\gl}{1-\gl} -\gd_n
\right)^{2^n}\right)\, , 
\end{equation}
where 
\begin{equation}
\gd_n:=\gd
 \left(  e^{\frac{s}{w ^n}} -1\right)\stackrel{n \to \infty} \sim c_\gl \frac{s}{w ^n}\, .
 \end{equation}
 It is now a matter of applying Lemma~\ref{th:ell}: the net result is 
 the following representation
for the generating function of the Harris random variable:
\begin{equation}
\label{eq:genHrv}
\psi(s)\, =\, q^{-1}
\left(\mathtt G \left(
\exp\left(
- (s \mathtt c) ^{\gamma}  \ga( \log (s\mathtt c) ) 
\right)
\right)
\right)\, ,
\end{equation}
where we recall that $\mathtt c=p_2/\gl$. 
Since $q^{-1} \circ \mathtt G (x)$ for $x \searrow 0$
behaves like a constant times $1/x$, one readily recovers 
\eqref{eq:Harris} (let us remark that in fact  \eqref{eq:genHrv} is  equivalent to \eqref{eq:Hrep}).
\end{rem}

\section{On quantitative estimates}
\label{sec:quantit}

\subsection{Estimating $\gO(\cdot)$}
A detailed numerical approach to $\gO(\cdot)$ can be found in \cite{cf:BN}
for $f(x)=\gl x +(1-\gl) x^2$, for $\gl= 0.1, 0.2,\ldots, 0.9$, but without explicit error bounds. 
The (two) methods they employ however do allow 
an explicit control of the error, with a non-trivial amount of work. Notably, by using what they call
{\sl B\"ottcher method}, which is based on \eqref{eq:Hrep}, one uses \eqref{eq:F.1} for $\tf(\cdot)$, for which
it is straightforward to control the error when one truncates the series,
and one can easily set up an iterative procedure to get the Taylor coefficients of $\psi(\cdot)$
with a control on the remainder by fixed point arguments. Note that it is sufficient to control 
the error for $s\in [c, cw]$, for a conveniently chosen $c$. Similar ideas are developed in detail just below for
$\go(\cdot)$, on which we focus, and we recover $\gO(\cdot)$ from $\go(\cdot)$.

\subsection{Estimating $\go(\cdot)$}
By recalling \eqref{eq:CH-rep} and by the  properties stated right after that formula we see that
\begin{equation}
\label{eq:get-om}
\go(\log s)\, =\, s^{-b} \mathtt{g}^{-1}
\left( \gp(\exp(-s))- \frac{1-\gl}\gl \right)\, ,
\end{equation}
where (see \cite{cf:CH})
\medskip
\begin{itemize}
\item $\mathtt{g}(\cdot)$ is the unique solution to
\begin{equation}
\label{eq:g}
\mathtt{g}(y)\, =\, A(\mathtt{g}(y/w))\, , \ \ \text{ with } \ \ A (y)\, :=\, wy+ \gl y^2\, , 
\ 
\mathtt{g}(0)\, =\, 0\, , \ \mathtt{g}'(0)\, =\, 1\, ;
\end{equation}
\item  $\gp(\cdot)$ is the unique solution of 
\begin{equation}
\label{eq:Pit}
\gp(y^2)\, =\, P(\gp(y))\, , \ \ \text{ with } \ \ P(x)\, =\, \gl x(1+x)\, ,
\end{equation} 
under the condition that $\gp(\cdot)$ is analytic in the  punctured  unit disc 
and that $\gp(z) \sim 1/ (\gl z)$ for $z$ going to zero.
\end{itemize}
\medskip

We recall the relation
\begin{equation}
\label{eq:gtog}
g(\cdot)\, =\, 1+ \frac{p_2}{\gl} \mathtt g (\cdot)\, ,
\end{equation}
and $g(\cdot)$ is the function appearing in Theorem~\ref{th:Bottcher}.
\medskip

In principle, it is now a matter of exploiting \eqref{eq:g} and \eqref{eq:Pit}
to get enough terms in the power series of $\mathtt g$ and $\gp$ and of controlling the remainder,
to obtain a sufficiently good approximation of $\go$
on $[x, x+\log 2]$, for some $x$, from which we can extract the
Fourier coefficients of $\go$ by controlled numerical integration. While in principle
this whole procedure is straightforward, in practice it is quite non-trivial given the 
fast decay of the coefficients and the fact that even the first coefficients are extremely small.

\medskip

\begin{rem}
\label{rem:add}
\rm 
Once $\go(\cdot)$ is precisely estimated over an interval of length $\log 2$ -- the period --
it is of course known with the same precision over $\bbR$. The plot of the Julia set in \ref{fig:1} is obtained by following the
same principle: we will not perform explicit estimates for $\go(\cdot)$ on $\bbC$ and we content ourselves with remarking the somewhat surprising 
precision of such a procedure, see Fig.~\ref{fig:1} and its caption. The graph in  Fig.~\ref{fig:1}
has been obtained by keeping 250 terms in both the series for $\mathtt g$ and $\gp$, and by performing three times the {\sl backward iteration procedure} that we explain just below.  
\end{rem}

\subsection{Approximating $\mathtt g $ and $\mathtt g ^{-1}$}
From \eqref{eq:g} we see that the power series for $\mathtt g$ has positive coefficients. One
can obtain the first $n$ coefficients by setting $g_{(n)}(x):= x +g_2x^2 + \ldots+ g_nx^n$, by computing the polynomial $g_{(n)}(wx)-A(g_{(n)}(x))$ and by setting to zero the coefficients of the terms of degree smaller than
$n+1$. This determines $g_2,\ldots, g_n$. Then we define $q$ via $\mathtt g(x)= g_{(n)}(x)+y^{n+1}q(x)$
and use \eqref{eq:g} to obtain
\begin{equation}
q(x)\, =\, Q(x) + p(x)q(x/w)+ \frac{\gl}{w^{2(n+1)}}x^{n+1} \left( q(x/w)\right)^2\, ,
\end{equation}
where $Q$ is a polynomial of degree $n-1$ and $Q(0)\neq 0$, $p(x):=w^{-n}+ 2 \gl w^{-n-1} g_{(n)}(x/w)$.
With the notation $\Vert q \Vert_\gep:= \max_{x \in [0,\gep]} \vert q(x)\vert$ we directly obtain that 
if we can exhibit $a>1$ and $\gep>0$ such that 
\begin{equation}
c_\gep\, :=\, 1- \Vert p \Vert _\gep -  a\frac{\gl}{w^{2(n+1)}}\gep ^{n+1} \vert Q(0) \vert >0\, 
\ \ \ \text{ and }\ \ \ 
\Vert Q\Vert _\gep  \, \le a c_\gep \vert Q(0) \vert \, , 
\end{equation}
then $\vert q(x)\vert \le a \vert Q(0) \vert$ for $\vert x \vert \le \gep$.
Therefore, by the positivity of the series coefficients, we have
\begin{equation}
g_{(n)}(x) \, \le \, 
\mathtt g (x)  \, \le \, g_{(n)}(x) +  a \vert Q(0) \vert x^{n+1}\, ,
\end{equation}
 for $x \in [0, \gep]$. By inverting the two polynomials that bound $\mathtt g$ from below and above,
 and by taking the Taylor expansion to order $n$ of these two expressions
 one directly recovers the power series for 
 $\mathtt g^{-1}$ truncated at $n$, with an explicit control on the rest.
  
 A performing way to improve this approximation of $\mathtt g ^{-1}$ 
is the following:
  from 
\eqref{eq:g} we obtain
\begin{equation}
\mathtt{g}^{-1} (y) \, =\, w\, \mathtt{g}^{-1}\left( A^{-1}(y)\right)\, ,
\end{equation}
where
\begin{equation}
A^{-1}(y) \, =\, \frac{\sqrt{4y \gl + w^2} \, -w}{2\gl}\, .
\end{equation}
$A^{-1}$ is  concave, with slope $1/w$ at the origin and from this we get
that if $\mathtt{g}^{-1} (y)= p(y)+r_0(y)$, with $p(\cdot)\ge0$ and $r_0$ a remainder like above, we get
\begin{equation}
\left \vert \mathtt{g}^{-1} (y) -w p\left(A^{-1}(y)\right) \right\vert \, \le \, w \left\vert r_0\left(A^{-1}(y)\right) 
\right\vert \, \le C w \left( \frac{y}{w}\right)^k\, .
\end{equation}
Note that 
the new remainder is better both because it improves by a
factor $w^{1-k}$
the estimate in the interval in which we have the estimate for $r_0$ (with no
a priori condition on the argument of $r_0$) and because it yields  an explicit estimate 
on an interval that is $w$ times larger. 
We sum up this argument:
\medskip

\begin{fact}
If $\mathtt{g}^{-1} = p+r_0$, with $p(x)\ge0$, $\vert r_0 (x)\vert \le C x^k$ for
$x\in [0, x_0]$ and if we set
\begin{equation}
\label{eq:fact1}
r_n(y)\, :=\, \mathtt{g}^{-1}  (y) -w^n p\left( (A^{-1})^{\circ n} (y) \right)\, , 
\end{equation}
then $\vert r_n(y) \vert \le w^n \vert r_0 ((A^{-1})^{\circ n}(y))\vert \le 
C y^k/ w^{(k-1)n}$, for $ y \in [0, w^n x_0]$ (actually, even on a much larger set,
since $A^{\circ n}(y)$ becomes much larger than $yw^n$ for $n$ large). 
\end{fact}

 \subsection{Approximating $\gp$}
 One starts by guessing the (Laurent) series coefficients for $\gp$, by using \eqref{eq:Pit}, and 
 it is not difficult to see that
 \begin{equation}
 \gp(x) \, =\, \frac 1{\gl x} - \frac 12 + p (x) + y^{2n+1}r(y)\, ,
 \end{equation} 
 where $r$ is analytic and $p(x)$ is a polynomial of degree $2n-1$
 that contains only odd powers of $x$. Again by using \eqref{eq:Pit} 
 we extract an equation for $r$:
 \begin{equation}
 r(x)p_1(x)+p_0(x)- \gl x^{2n+2} r_n(x)^2+ r_n(x^2) x^{2n+2}\, =\, 0\,,
 \end{equation}
 where $p_1(x)$ is a polynomial of degree $2n$, containing only even powers, with $p_1(0)=-2$;
 $p_0(x)$ contains also only even powers and it is a polynomial of degree $2n-2$.
 Then one can show that if one can exhibit $a>1$ and $\gep>0$ such that
 \begin{equation}
 C_\gep \, :=\, 1- \left\Vert \frac 1{p_1} \right\Vert_\gep  \gep ^{2n+2}
 \left( a \gl \left \vert \frac{p_0(0)}{p_1(0)} \right\vert +1 \right) \, >\,  0
 \ \ \ \text{ and } \ \ \ 
 \left \Vert \frac{p_0}{p_1} \right\Vert_\gep \, \le \, a C_\gep \left\vert \frac{p_0(0)}{p_1(0)} \right\vert\, ,
 \end{equation}
 then 
\begin{equation}
 \left \vert 
 \gp(x) - \left(\frac 1{\gl x} - \frac 12 + p (x)\right) \right \vert  \, \le \,  
 \frac a 2 \vert p_0(0) \vert  
  y^{2n+1}\, ,
 \end{equation} 
 for $(0, \gep]$.

The approximation of $\gp$ can be greatly
improved by backward iteration, as for $\mathtt g^{-1}$: from \eqref{eq:Pit} it follows that
\begin{equation}
\label{eq:gpimp}
\gp(y)\, =\, Q(\gp(y^2))\, , \ \ \ \text{ where } \ \ Q(x)\, =\, \frac{\sqrt{1+ \frac{4x}{\gl}}\, -1}2\, .
\end{equation}
Note that $Q(\cdot)$ is concave, $Q'(0)=1/\gl$ and $Q(x)=x$ implies $x=0$
or $x=(1-\gl)/\gl$. So if
$\gp= \tilde \gp + r_0$, with $\tilde \gp(x)\ge 0$ and $\vert r_0(x)\vert \le C  x ^k$
for $x \in (0, x_0]$, then 
\begin{equation}
\label{eq:ests1}
\left \vert \gp (y) - Q(\tilde \gp (y^2))\right \vert \, \le \, \frac 1 \gl \vert r_0(y^2)\vert\,  \le \, \frac C \gl  y ^{2k}\,,
\end{equation}
for $y\in (0, \sqrt{x_0}]$.
This can also be seen by using $\max_{x \ge 0}\vert Q'(x)\vert = 1/\gl$. 
Note that \eqref{eq:ests1} implies that $Q(\tilde \gp (y^2))$
is better that $ \tilde \gp(y)$ in approximating $\gp(y)$ in the sense that the new remainder
($r_1(y):= \gp(y)-Q(\tilde \gp (y^2))$)
is smaller than the previous one if  $y$ is in the
original interval, that is $ y\in (0, x_0]$,  if $x_0< \gl^{1/k}$, 
and one has an explicit estimate for $r_1$ in a larger interval.
To sum up:

\begin{fact} If 
$\gp= \tilde \gp + r_0$, with $\tilde \gp(x)\ge 0$ and $\vert r_0(x)\vert \le C  x^k$
for $x \in (0, x_0]$, then if we set
\begin{equation}
\label{eq:fact2}
r_n(y):= \gp(y)-Q^{\circ n} \left( \tilde \gp \left(y^{2^n}\right)\right)\, ,
\end{equation}
we have $\vert r_n(y) \vert \le \frac 1{\gl^n} \vert r_0 (y^{2^n}) \vert\le C y ^{2^n k}/\gl^n$,
for $y\in [0, x_0^{1/2^n}]$.
\end{fact}
\medskip

\subsection{Example of computation}
By applying the arguments of the previous subsection for $\gl=2-w=1/5$ we have
\begin{multline}
\mathtt{g}(x)\, =\, 
x + \frac{(2 - w) x^2}{(w-1) w} + \frac{
 2 (2 - w)^2 x^3}{(w-1)^2 w^2 (1 + w)} +\frac{ ( 2-w)^3 (5 + 
    w) x^4}{( w-1)^3 w^3 (1 + w) (1 + w + w^2)} \\ + \frac{
 2 (w-2)^4 (7 + w (3 + 2 w)) x^5}{( w-1)^4 w^4 (1 + w)^2 (1 + 
    w^2) (1 + w + w^2)} +r_5(x)\, ,
\end{multline}
with $\vert r_5(x) \vert \le 5 \vert x \vert ^6 /10^6$ for $\vert x\vert \le 2.5$.
From this we extract
\begin{multline}
\mathtt{g}^{-1}(x)\, =\, 
x-\frac{\gl x^2}{(-1+w) w}+
\frac{2 \gl^2 x^3}{(-1+w)^2 w (1+w)}-\frac{\gl^3 \left(1+5 w^2\right) x^4}
{(-1+w)^3 w^2 (1+w) \left(1+w+w^2\right)}
\\
+\frac{2 \gl^4 \left(3+2 w+7 w^3\right) x^5}{(-1+w)^4 w (1+w)^2 \left(1+w^2\right) \left(1+w+w^2\right)} + r_5^- (x)\, ,
\end{multline}
with $\vert r_5^- (x) \vert \le 5 \vert x\vert ^6/10^4$ for $\vert x \vert \le 2$.
Moreover 
\begin{multline}
\gp(x)\, =\, \frac{1}{\gl x}
 -\frac{1}{2}+\frac{1}{8} (-2+\gl) x+\frac{1}{128} \left(-16+4 \gl+4
\gl^2-\gl^3\right) x^3\\ +
\frac{(-2+\gl) \gl \left(16-4 \gl-4 \gl^2+\gl^3\right) x^5}{1024}+ R_6(x)\, ,
\end{multline}
where $\vert R_6(x)\vert \le x^7/10$ for $x \in (0, 9/10]$.

\begin{figure}[hlt]
\begin{center}
\leavevmode
\epsfxsize =14 cm
\epsfbox{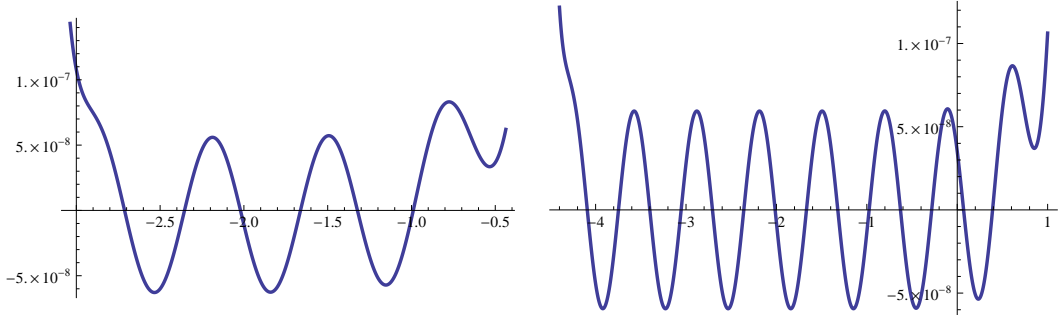}
\end{center}
\label{fig:3}
\caption{
The plots of approximations of $\go(\cdot)-\overline{\go}$, cf. \eqref{eq:overgo}, for $\gl=1/5$
are obtained by using the expansion of $\mathtt g^{-1}$ to $5^{\textrm{th}}$ order and $\gp$ to  $6^{\textrm{th}}$ order, and by 
applying five times, on the left, and seven times, on the right, the backward iteration procedure.
The case on the right is used for the quantitative estimates, and the interval $[-2-\log 2, -2]$, in which we have good error estimates,  is chosen.}
\end{figure}

Aiming at obtaining a good approximation of $\go$ we choose
$s \in [1/(2e^2), 1/e^2]
$, so $\exp(-s) \in [0.873\ldots, 0.934\ldots]$.
For this interval of values we  obtain $\gp(\exp(-s))-(1-\gl)/\gl \in [0.48\ldots, 0.92\ldots]$
with an error of at most $(0.93)^{7 \times 2^{n_1}  }/(10 (0.2)^{n_1}$,
where $n_1$ is the number of backward iterations employed, {\sl cf.} \eqref{eq:fact2}.
Similarly, the error on $\mathtt g ^{-1}$ in the range of values under 
consideration is uniformly bounded by $(5/10^4) (0.93)^6/(1.8)^{5n_2}$.
Choosing $n_1=7$ makes the error on $\gp$ smaller than $10^{-23}$
and setting also $n_2=7$ leads to an error on $\go$ that is uniformly bounded 
by $5\times10^{-13}$. This allows to estimate the {\sl average} value of $\go$
and the first Fourier coefficient with a precision of at least $10^{-11}$:
$\go(x)= \overline{\go}+ g_1 \sin (2\pi (x-x_0)/ \log 2)+\ldots$
\begin{equation}
\label{eq:overgo}
\overline{\go} := \frac 1{\log 2}\int_{-2-\log (2)}^{-2 } \go (y) \dd y \, \approx\, 
4.45140273002
\end{equation}
and $g_1\approx 5.938 \times 10^{-8}$.

At this point if one wants to recover $\gO(\cdot)$ one has to perform the inversion step in
Theorem~\ref{th:Bottcher}(2). It is not difficult to realize by considering
\eqref{eq:forinv} that, given the small size of the oscillations,  a good approximation of $\ga(y)$ (defined in Theorem~\ref{th:Bottcher}(2))
is $1/\go^\gamma( \gamma  y -\gamma \log \overline \go)$). Of course a control of the error requires an attentive (but elementary) analysis. 
By performing explicitly the case under consideration and choosing for example  $p_1=0$, $p_2=w/2$ and $p_0=1-p_2$
we obtain 
$\overline \gO= 1.01288677326$ and the first Fourier coefficient is $1.59\times 10^{-8}$.

\section*{Acknowledgments} 
G.~G. thanks Bernard Derrida and Mathieu Merle for enlightening discussions. 
The research of O.~C. was supported in part by NSF DMS grant 1108794.
G.~G.
acknowledges the support of ANR (grant SHEPI), Univ. Paris Diderot (project SCHePS) and the 
Petronio Fellowship Fund at IAS (Princeton).

\end{document}